\documentclass[prl,amsmath,amsfonts,amssymb,twocolumn,showpacs]{revtex4}
\usepackage{hyperref}
\usepackage{graphicx}

\def\tr{\mbox{tr}}
\newcommand \norm[1]{\|#1\|}

\begin{document}

\title{Generalized Limits for Single-Parameter Quantum Estimation}
\author{Sergio Boixo}
\author{Steven T. Flammia}
\author{Carlton M. Caves}
\author{JM Geremia}
\affiliation{Department of Physics and Astronomy, University of New Mexico,
Albuquerque, New Mexico 87131}

\date{\today}

\begin{abstract}
We develop generalized bounds for quantum single-parameter estimation
problems for which the coupling to the parameter is described by intrinsic
multi-system interactions.  For a Hamiltonian with $k$-system
parameter-sensitive terms, the quantum limit scales as $1/N^k$ where $N$
is the number of systems.  These quantum limits remain valid when the
Hamiltonian is augmented by any parameter-independent interaction among
the systems and when adaptive measurements via parameter-independent
coupling to ancillas are allowed.
\end{abstract}

\maketitle

Many problems that lie at the interface between physics and information
science can be addressed using techniques from parameter estimation
theory.   Precision metrology, timekeeping, and communication offer
prominent examples; the parameter of interest might be the strength of an
external field, the evolved phase of a clock, or a communication symbol.
Fundamentally, single-parameter estimation is a quantum-mechanical
problem: one must infer the value of a coupling constant $\gamma$ in the
Hamiltonian $H_\gamma=\hbar\gamma h_0$ of a probe system by observing the
evolution of the probe due to
$H_\gamma$~\cite{Helstrom1976,Holevo1982,Braunstein1994,Braunstein1996,Giovannetti2006}.
We take $\gamma$ to have units of frequency, thus making $h_0$ a
dimensionless coupling Hamiltonian.

Quantum mechanics places limits on the precision with which $\gamma$ can
be determined.   It is now well established, via the quantum Cram\'er-Rao
bound~\cite{Helstrom1976,Holevo1982,Braunstein1994,Braunstein1996}, that
the optimal uncertainty in any single-parameter quantum estimation
procedure is $\delta\gamma\sim1/\sqrt{\nu}\,t\,\Delta h_0$, where $\nu$ is
the number of independent probes used, $t$ is the evolution time of each
probe, and $\Delta h_0$ is the standard deviation (uncertainty) of
$h_0$~\cite{Braunstein1994,Braunstein1996}.  The $1/\sqrt{\nu}$ dependence
is the standard statistical improvement with number of trials; generally,
for nonGaussian statistics, the sensitivity $1/\sqrt{\nu}\,t\,\Delta h_0$
can only be attained asymptotically for a large number of trials. Besides
increasing the number of trials, there are two other obvious ways to
improve the sensitivity: (i)~the probe can be allowed to evolve under
$H_\gamma$ for a longer time $t$; (ii)~the quantum state of the probe can
be chosen to maximize the deviation $\Delta h_0$.  In all practical
settings, decoherence or other processes limit the useful interaction
time; in addition, temporal fluctuations in $\gamma$ often prevent the
evolution time from being arbitrarily extended.  For a given parameter
estimation problem, $h_0$ is fixed, as is its maximum deviation.

Quantum mechanics does, however, provide another opportunity: gathering
$N$ probe systems into a single probe, which is prepared in an appropriate
entangled state; if $\Delta h_0$ for the entangled state increases faster
than $\sqrt N$, the sensitivity improves, provided there is still a
sufficient number of probes to reach the asymptotic regime in number of
trials.  This Letter focuses on how $\delta\gamma$ scales with $N$, the
number of systems in a probe.  Thus we work throughout with bounds on the
sensitivity of a single probe, remembering that the bounds can only be
achieved by averaging over many probes, but preferring not to muddy the
discussion by carrying along the $1/\sqrt\nu$ dependence on the number of
probes.

For the $N$ systems in a probe, it has been traditional to consider
Hamiltonians of the form
\begin{equation} \label{eq:Hgamma}
    H_\gamma = \hbar\gamma h_0\;,
    \qquad
    h_0=\sum_{j=1}^N h_j\;,
\end{equation}
where the $h_j$'s are single-system dimensionless coupling Hamiltonians,
assumed to be identical.  Restriction to Hamiltonians that are separable
and invariant under particle exchange, as in Eq.~(\ref{eq:Hgamma}), is
physically motivated: in metrology it is generally desirable to make the
coupling to the parameter homogeneous, and multi-body effects are
typically undesirable because they are less well characterized.  In atomic
clocks, for instance, much experimental effort is directed toward
achieving a Hamiltonian of the form~(\ref{eq:Hgamma}).  In many
cases, even the measurements performed on the probe are unable to
distinguish between individual constituents.

To determine how the optimal parameter uncertainty scales with $N$, one
maximizes the deviation $\Delta h_0$ over joint states of the probe
systems.  If entanglement is not allowed, the probe systems can be
regarded themselves as independent probes; in this case, $\Delta h_0$
scales as $\sqrt{N}$, producing the so-called shot-noise limit found in
precision magnetometry, gravimetry, and timekeeping~\cite{Wineland1994}.
When entanglement is allowed, however, one can choose the initial probe
state to be the ``cat state,''
\begin{equation}\label{eq:cat}
{1\over\sqrt2}\Bigl(
|\lambda_M,\ldots,\lambda_M\rangle
+|\lambda_m,\ldots,\lambda_m\rangle\Bigl)\;,
\end{equation}
where for system~$j$, $|\lambda_M\rangle$ ($|\lambda_m\rangle$) is the
eigenstate of $h_j$ with maximum (minimum) eigenvalue $\lambda_M$
($\lambda_m$).  This yields a deviation $\Delta h_0 =
N(\lambda_M-\lambda_m)/2$ that scales linearly in
$N$~\cite{Braunstein1994,Giovannetti2006}, a scaling known as the
Heisenberg limit.  Evolution under $H_\gamma$ for time $t$ introduces a
relative phase $e^{i\phi(t)}$ into the cat state, with $\phi(t)=\gamma
tN(\lambda_M-\lambda_m)$, and leaves $\Delta h_0$ unchanged; the
Heisenberg limit can be attained (asymptotically for many trials) by
measuring on each probe system an observable two of whose eigenvectors
are $|\pm\rangle=(|\lambda_M\rangle\pm|\lambda_m\rangle)/\sqrt2$.

The Heisenberg limit is not general since there are physical systems of
interest for parameter estimation where a coupling Hamiltonian of the
form~(\ref{eq:Hgamma}) is overly restrictive. In particular, some
condensed and even quantum-optical systems exhibit nonlinear collective
effects due to multi-body or tensor-field
interactions~\cite{Botet1982,Cirac1998,You2006}.  In such systems,
multi-body terms in the Hamiltonian can couple to metrologically relevant
parameters. In this Letter we generalize single-parameter quantum
estimation to intrinsic many-body interactions, which surpass the
conventional Heisenberg limit. Our work is largely inspired by a recent
paper of Roy and Braunstein~\cite{Roy2006}, which claimed an exponential
scaling for a collection of $N$ qubits with a particular Hamiltonian.  We
argue below that the exponential scaling is unphysical.

We turn now to showing that Hamiltonians with intrinsic $k$-body terms
generate a family of parameter estimation problems, characterized by $k$,
where the quantum limit scales as $1/N^k$.  For this purpose, we consider
Hamiltonians of the form
\begin{equation}\label{eq:ham2}
    H_\gamma(t) = \hbar\gamma h_0 + \tilde H(t)\;,
    \quad
    h_0 = \sum_{\{j_1,\ldots,j_k\}} h_{j_1, \ldots,j_k}^{(k)}\;,
\end{equation}
where $h_0$ is the dimensionless Hamiltonian that describes coupling to
the parameter.  The auxiliary Hamiltonian $\tilde H(t)$ is discussed
below. In $h_0$, $k$ denotes the degree of multi-body coupling, with the
sum running over all subsets of $k$ systems.  We could also include
couplings of different degrees up to a maximum degree, but since the
maximum degree dominates the sensitivity scaling, we stick with a single
degree $k$ in the following.  We assume that the $k$-body coupling
$h^{(k)}$ is symmetric under exchange of probe systems. Moreover, we
assume that $k$ and $h^{(k)}$ are independent of the number of probe
systems.  We make this latter assumption, that $h_0$ is an intensive
property of the probe, because we want to consider a particular kind of
coupling to the parameter which remains unchanged as $N$ changes. For real
physical systems, the symmetry and intensive assumptions will hold only
approximately and only over some range of values of $N$.

\begin{figure*}[t]
\begin{center}
\includegraphics{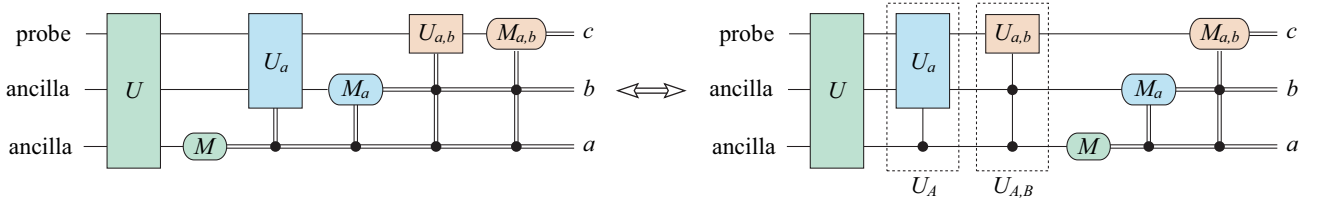}
\end{center}
\vspace{-4mm}
\caption{Quantum-circuit diagrams illustrating the principle of deferred
measurement.  In the circuit on the left, a measurement $M$ on the lower
ancilla yields result $a$; this result controls a subsequent unitary
$U_a$, applied to the probe and the upper ancilla, and determines a
conditional measurement $M_a$ on the upper ancilla, which has result~$b$.
The two measurement results then control a unitary $U_{a,b}$ applied to
the probe and a conditional measurement $M_{a,b}$ on the probe.  The
left-hand circuit is equivalent to the circuit on the right, in which the
controls are applied coherently (boxed gates $U_A$ and $U_{A,B}$) and the
measurements, deferred to the end of the circuit, tell one which unitary
was applied.  Without loss of generality, we can assume the measurements
are described by orthogonal projectors $P_a$, $P_{b|a}$, and $P_{c|a,b}$,
because any generalized measurement can be modeled by a projection-valued
measurement on an extended system.  The unitary transformations in the
left-hand circuit, $U_a$ and $U_{a,b}$, are evolution operators generated
by the Hamiltonians $\hbar\gamma h_0+\tilde H_a(t)$ and $\hbar\gamma
h_0+\tilde H_{a,b}(t)$, whereas the corresponding coherent controlled
unitaries in the circuit on the right, $U_A$ and $U_{A,B}$, are generated
by the Hamiltonians $\hbar\gamma h_0+\sum_a\tilde H_a(t)\otimes P_a$ and
$\hbar\gamma h_0+\sum_{a,b}\tilde H_{a,b}(t)\otimes P_{b|a}P_a$.  It is
easy to verify from the evolution equations that the controlled unitaries
in the right-hand circuit are given by $U_A=\sum_a U_a\otimes P_a$ and
$U_{A,B}=\sum_{a,b}U_{a,b}\otimes P_{b|a}P_a$.  Thus the principle of
deferred measurement can be rendered algebraically in the following way:
if we use the left-hand circuit, the probability for obtaining results
$a$, $b$, and $c$ takes the form $\tr(C_{a,b,c}\rho_0C^\dagger_{a,b,c})$,
where $\rho_0$ is the initial state of the probe and ancillas and
$C_{a,b,c}=P_{c|a,b}U_{a,b}P_{b|a}U_aP_aU$; pulling the measurement
projectors to the left in $C_{a,b,c}$ changes the unitaries to the
corresponding coherent controlled operations, i.e.,
$C_{a,b,c}=P_{c|a,b}P_{b|a}P_aU_{A,B}U_AU$, which gives the form of the
probability obtained from the right-hand circuit.} \label{F:circuit}
\end{figure*}

The auxiliary Hamiltonian $\tilde H(t)$ includes all parameter-independent
contributions to $H_\gamma$.  For example, it includes the free
Hamiltonians of the probe systems and any parameter-independent
interactions among them.  In addition, we can introduce an undetermined
number of ancillas and let $\tilde H$ include the couplings of the
ancillas to the probe systems and any couplings among the ancillas.
Measurements on the ancillas can be included as part of an overall final
measurement on the probe-ancilla system; since the Cram\'er-Rao bound that
underlies our analysis holds for all possible measurements and ways of
estimating $\gamma$ from the measurement results, the bounds we derive
hold for arbitrary measurements on the ancillas.  This conclusion applies
even to measurements on the ancillas that are carried out during the
evolution time and whose results are used to condition measurements on
other ancillas or to control the coupling of other ancillas to the probe.
By the principle of deferred measurement~\cite{Nielsen2000}, which is
illustrated in Fig.~\ref{F:circuit}, all such measurements can be shuffled
to the end of the evolution time by making appropriate adjustments to
$\tilde H$.

Now let $\rho_0$ be the initial state of the probe and any ancillas.
After a time~$t$, the state evolves to
$\rho_\gamma(t)=U_\gamma(t)\rho_0U^\dagger_\gamma(t)$, where the evolution
operator is generated by the Hamiltonian~(\ref{eq:ham2}):
\begin{equation}\label{eq:generator}
i\hbar{\partial U_\gamma(t) \over \partial t}=H_\gamma(t)U_\gamma(t)\;.
\end{equation}
At time $t$, measurements are made on the probe and ancillas, the results
of which are used to make an estimate $\gamma_{\rm est}$ of the parameter.
The appropriate statistical measure of the estimate's precision is the
units-corrected mean-square deviation of $\gamma_{\rm est}$ from
$\gamma$~\cite{Braunstein1994,Braunstein1996}:
\begin{equation}
\delta^2\gamma=
\biggl\langle\biggl(
{\gamma_{\rm est}\over|d\langle\gamma_{\rm est}\rangle/d\gamma|}
-\gamma
\biggr)^{\!2}
\biggr\rangle\;.
\end{equation}
Here and below expectation values are evaluated with respect to
$\rho_\gamma(t)$.

The quantum Cram\'er-Rao bound states
that~\cite{Helstrom1976,Holevo1982,Braunstein1994,Braunstein1996}
\begin{equation}
\delta^2\gamma\ge{1\over\mathcal{I}_\gamma(t)}\;,
\quad
\mathcal{I}_\gamma(t)
=\tr{\bigl(\rho_\gamma(t)\mathfrak L_{\gamma}^2(t)\bigr)}
=\langle\mathfrak{L}^2_\gamma(t)\rangle\;,
\end{equation}
where $\mathcal{I}_\gamma(t)$ is the quantum Fisher information. The
Hermitian operator $\mathfrak L_\gamma(t)$, called the symmetric
logarithmic derivative, is defined (implicitly) by
\begin{equation}\label{eq:geneq}
\frac{1}{2}(\mathfrak L_\gamma \rho_\gamma + \rho_\gamma \mathfrak L_\gamma)
=\frac{\partial \rho_\gamma}{\partial \gamma}=
-i[K_\gamma,\rho_\gamma]\;.
\end{equation}
Here
\begin{equation}
K_\gamma(t)=i\frac{\partial U_\gamma(t)}{\partial \gamma} U^\dag_\gamma(t)
\end{equation}
is the Hermitian generator of displacements in $\gamma$.  If there is
no auxiliary Hamiltonian, $K_\gamma(t)=th_0$.

For pure states, differentiating $\rho_\gamma=\rho_\gamma^2$ shows that
\begin{equation}\label{E:sld}
\mathfrak L_{\gamma}(t)
=2\frac{\partial \rho_\gamma(t)}{\partial\gamma}
=-2i[K_\gamma(t),\rho_\gamma(t)]\;.
\end{equation}
Then the Fisher information reduces to a multiple of the variance of
$K_\gamma(t)$:
\begin{equation}
\mathcal I_\gamma(t)=4\bigl(\langle K_\gamma^2(t)\rangle-\langle K_\gamma(t)\rangle^2\bigr)
=4\Delta^2 K_\gamma(t)\;.
\end{equation}
For mixed states, the variance provides an upper bound on the Fisher
information, instead of equality~\cite{Braunstein1996}.

We define the operator semi-norm $\norm{H}$ of a Hermitian operator $H$ as
$\norm{H}=M_H-m_H$, where $M_H$ ($m_H$) is the maximum (minimum)
eigenvalue of $H$.  This semi-norm is invariant under unitary
transformations and obeys the triangle inequality, i.e.,
$\norm{H+K}\le\norm{H}+\norm{K}$~\cite{note1}. The importance of the
semi-norm is that its square provides an upper bound on the variance,
i.e., $\Delta^2 H\le\norm{H}^2/4$~\cite{note2}. The maximum variance is
achieved by pure states $(|M_H\rangle+e^{i\phi}|m_H\rangle)/\sqrt2$.

We can now summarize the chain of inequalities satisfied by the estimation
precision,
\begin{equation}
\label{E:ineqaulities}
    {1\over\delta\gamma}
    \le\sqrt{\mathcal I_\gamma(t)}
    \le2\Delta K_\gamma(t)
    \le\norm{K_\gamma(t)}\;,
\end{equation}
leaving us with the final task of bounding the semi-norm of $K_\gamma(t)$
for the dynamics of Eqs.~(\ref{eq:ham2}) and (\ref{eq:generator}).  To do so, we
define a new Hermitian operator,
\begin{equation}
F_\gamma(t) = U^\dag_\gamma(t) K_\gamma(t) U_\gamma(t)
= iU^\dag_\gamma(t)\frac{\partial U_\gamma(t)}{\partial \gamma}\;,
\end{equation}
which satisfies the evolution equation, $\partial F_\gamma(t)/\partial t =
U^\dag_\gamma(t)h_0U_\gamma(t)$, with initial condition
$F_\gamma(0)=iU^\dag_\gamma(0)(\partial U_\gamma(0)/\partial \gamma)=0$,
since $U_\gamma(0)=I$.  Straightforward integration provides
$F_\gamma(t)$, and conversion back to $K_\gamma(t)$ gives
\begin{equation}
K_\gamma(t) =
\int_0^t ds\,U_\gamma(t)U^\dag_\gamma(s) h_0 U_\gamma(s)U^\dag_\gamma(t)\;.
\end{equation}
The triangle inequality and the unitary invariance of the semi-norm imply
that
\begin{eqnarray}\label{eq:normK}
\norm{K_\gamma(t)}&\le&
\int_0^t ds\,\norm{U_\gamma(t)U^\dag_\gamma(s)h_0 U_\gamma(s)U^\dag_\gamma(t)}
\nonumber\\
&\le&t\norm{h_0}\;,
\end{eqnarray}
which gives us the desired bound on the sensitivity,
\begin{equation}
\delta\gamma\ge\frac{1}{t\norm{h_0}}\;.
\label{eq:trt}
\end{equation}

This bound on the estimation precision applies for any coupling
Hamiltonian $h_0$.  It shows that the optimal sensitivity is determined by
$h_0$---indeed, it is determined by the range of energies in $h_0$---and
cannot be improved by use of a parameter-independent auxiliary Hamiltonian
$\tilde H$ or of ancillas not coupled directly to the parameter, although
both of these might be used in physical settings to make the required
optimal measurement accessible \cite{Geremia2003}.

The result underlying the bound (\ref{eq:trt}) was obtained by Giovannetti
{\it et al.}~\cite{Giovannetti2006} for the case of discrete operations,
as opposed to continuous time evolution, and was used there to show that
multi-round protocols with single-system probes (and allowing for ancillas
and adaptive measurements) have the same optimal sensitivity as
single-round protocols with multi-system entangled probes.  We use the
bound in a different way, and our derivation shows directly that the
ultimate sensitivity cannot be improved when the auxiliary Hamiltonian
$\tilde H$ acts simultaneously with the coupling Hamiltonian $h_0$.

We now apply the bound~(\ref{eq:trt}) to draw physical conclusions about
the sensitivity scaling for the various forms of $h_0$.  For the
separable, symmetrically coupled Hamiltonian of Eq.~(\ref{eq:Hgamma}), we
recover the $1/N$ scaling of the Heisenberg limit.  For the symmetric
$k$-body coupling of Eq.~(\ref{eq:ham2}), the triangle inequality applied
to the semi-norm,
\begin{equation}\label{eq:normh0}
\norm{h_0}\le\sum_{\{j_1,\ldots,j_k\}}\norm{h^{(k)}_{j_1,\ldots,j_k}}=
{N\choose k}\norm{h^{(k)}}\sim {N^k\over k!}\norm{h^{(k)}}\;,
\end{equation}
gives a sensitivity limit that scales as $1/N^k$.

An important special case occurs when $\tilde H(t)=0$, so that
$K_\gamma(t)=th_0$, the $k$-body coupling terms in $h_0$ are products of
single-system operators, i.e., $h^{(k)}_{j_1,\ldots,j_k}=h_{j_1}\cdots
h_{j_k}$, and the single-system operators have nonnegative eigenvalues.
Then the inequalities in Eqs.~(\ref{eq:normK}) and (\ref{eq:normh0})
become equalities, and an initial cat state~(\ref{eq:cat}) attains the
maximum deviation, i.e.,
\begin{equation} \label{eq:kbound}
\Delta K_\gamma(t)={1\over2}\norm{K_\gamma(t)}={t\over2}\norm{h_0}=
{t\over2}{N\choose k}(\lambda_M^k-\lambda_m^k)\;.
\end{equation}
The brief discussion of attaining the Heisenberg limit ($k=1$), just after
Eq.~(\ref{eq:cat}), can be applied directly to achieving the sensitivity
limit for arbitrary $k$, except that the relative phase generalizes to
$\phi(t)=\gamma t{N\choose k}(\lambda_M^k-\lambda_m^k)$.

Our result can be used to analyze the recent paper by Roy and Braunstein
(RB) \cite{Roy2006}, which inspired the work we report here.  In our
notation, RB consider a system of $N$ qubits with dimensionless coupling
Hamiltonian
\begin{equation}
h_0={1\over2}(\Sigma_++\Sigma_-)\;,\qquad \Sigma_\pm=\prod_{j=1}^N(X_j\pm iY_j)\;,
\end{equation}
where $X_j$ and $Y_j$ are Pauli operators for the $j$th qubit.  When the
products are multiplied out, $h_0$ becomes a sum of $2^{N-1}$ commuting
Pauli products; it has maximum deviation $\Delta h_0=\norm{h_0}/2
=2^{N-1}$, which gives a quantum limit that scales exponentially in $N$.
RB suggest that their Hamiltonian describes the atomic transitions of $N$
atoms associated in a molecule, but the fundamental coupling in this case
is a separable sum, as in Eq.~(\ref{eq:Hgamma}), describing separate
transitions for each atom.  The RB coupling could arise as an effective
$N$th-order process, but it would not be justified to neglect processes of
other orders.  To achieve the RB Hamiltonian as a fundamental interaction
would require coupling the atoms to a rank-$N$ tensor field, but in this
case, every value of $N$ would involve a different fundamental coupling.
One could scarcely claim to be estimating the same coupling constant as
$N$ changes if the fundamental interaction is changing.

The most realistic possibility for taking advantage of multi-body
couplings in parameter estimation will be for pairwise couplings ($k=2$).
Hamiltonians with symmetrically parameterized two-body terms arise
naturally in field-theoretic systems, such as quantum degenerate gasses,
superconductors, and atomic ensembles coupled to a common electromagnetic
field mode.   Atom-atom interactions in a Bose-Einstein condensate (BEC)
\cite{Cornish2000} might offer a physically realistic approach to
surpassing the conventional Heisenberg limit, possibly even achieving
$1/N^2$ scaling.  We envisage situations where an external field modulates
the strength of the two-body scattering term in the second-quantized
condensate Hamiltonian.   Such a modulation occurs both for a
magnetically tuned Feschbach resonance and for density variations due to
gravitational gradients.

While exponential sensitivity improvements appear unphysical, more modest
quadratic or other polynomial improvements beyond the Heisenberg limit
could be essential for achieving the sensitivities required in the most
demanding precision measurements.   This work was supported in part by ONR
Grant No.~N00014-03-1-0426 and AFOSR Grant No.~FA9550-06-01-0178.  The
authors thank H.~Barnum, A.~Datta, S.~Merkel, R.~Raussendorf, and A.~Shaji
for helpful discussions.


\vspace{-1mm}
\begin{thebibliography}{15}
\expandafter\ifx\csname natexlab\endcsname\relax\def\natexlab#1{#1}\fi
\expandafter\ifx\csname bibnamefont\endcsname\relax
  \def\bibnamefont#1{#1}\fi
\expandafter\ifx\csname bibfnamefont\endcsname\relax
  \def\bibfnamefont#1{#1}\fi
\expandafter\ifx\csname citenamefont\endcsname\relax
  \def\citenamefont#1{#1}\fi
\expandafter\ifx\csname url\endcsname\relax
  \def\url#1{\texttt{#1}}\fi
\expandafter\ifx\csname urlprefix\endcsname\relax\def\urlprefix{URL }\fi
\providecommand{\bibinfo}[2]{#2}
\providecommand{\eprint}[2][]{\url{#2}}

\bibitem[{\citenamefont{Helstrom}(1976)}]{Helstrom1976}
\bibinfo{author}{\bibfnamefont{C.~W.} \bibnamefont{Helstrom}},
  \emph{\bibinfo{title}{Quantum detection and estimation theory}}, vol.
  \bibinfo{volume}{123} of \emph{\bibinfo{series}{Mathematics in science and
  engineering}} (\bibinfo{publisher}{Academic Press}, \bibinfo{address}{New
  York}, \bibinfo{year}{1976}), \bibinfo{edition}{1st} ed.

\bibitem[{\citenamefont{Holevo}(1982)}]{Holevo1982}
\bibinfo{author}{\bibfnamefont{A.~S.} \bibnamefont{Holevo}},
  \emph{\bibinfo{title}{Probabilistic and statistical aspects of quantum
  theory}}, vol.~\bibinfo{volume}{1} of \emph{\bibinfo{series}{North-Holland
  series in statistics and Probability theory}}
  (\bibinfo{publisher}{North-Holland}, \bibinfo{address}{Amsterdam},
  \bibinfo{year}{1982}), \bibinfo{edition}{1st} ed.

\bibitem[{\citenamefont{Braunstein and Caves}(1994)}]{Braunstein1994}
\bibinfo{author}{\bibfnamefont{S.~L.} \bibnamefont{Braunstein}}
  \bibnamefont{and} \bibinfo{author}{\bibfnamefont{C.~M.} \bibnamefont{Caves}},
  \bibinfo{journal}{Phys. Rev. Lett.} \textbf{\bibinfo{volume}{72}},
  \bibinfo{pages}{3439} (\bibinfo{year}{1994}).

\bibitem[{\citenamefont{Braunstein et~al.}(1996)\citenamefont{Braunstein,
  Caves, and Milburn}}]{Braunstein1996}
\bibinfo{author}{\bibfnamefont{S.~L.} \bibnamefont{Braunstein}},
  \bibinfo{author}{\bibfnamefont{C.~M.} \bibnamefont{Caves}}, \bibnamefont{and}
  \bibinfo{author}{\bibfnamefont{G.~J.} \bibnamefont{Milburn}},
  \bibinfo{journal}{Ann. Phys. (N.Y.)} \textbf{\bibinfo{volume}{247}},
  \bibinfo{pages}{135} (\bibinfo{year}{1996}).

\bibitem[{\citenamefont{Giovannetti et~al.}(2006)\citenamefont{Giovannetti,
  Lloyd, and Maccone}}]{Giovannetti2006}
\bibinfo{author}{\bibfnamefont{V.}~\bibnamefont{Giovannetti}},
  \bibinfo{author}{\bibfnamefont{S.}~\bibnamefont{Lloyd}}, \bibnamefont{and}
  \bibinfo{author}{\bibfnamefont{L.}~\bibnamefont{Maccone}},
  \bibinfo{journal}{Physical Review Letters} \textbf{\bibinfo{volume}{96}},
  \bibinfo{pages}{010401} (\bibinfo{year}{2006}).

\bibitem[{\citenamefont{Wineland et~al.}(1994)\citenamefont{Wineland,
  Bollinger, Itano, and Heinzen}}]{Wineland1994}
\bibinfo{author}{\bibfnamefont{D.~J.} \bibnamefont{Wineland}},
  \bibinfo{author}{\bibfnamefont{J.~J.} \bibnamefont{Bollinger}},
  \bibinfo{author}{\bibfnamefont{W.~M.} \bibnamefont{Itano}}, \bibnamefont{and}
  \bibinfo{author}{\bibfnamefont{D.~J.} \bibnamefont{Heinzen}},
  \bibinfo{journal}{Phys. Rev. A} \textbf{\bibinfo{volume}{50}},
  \bibinfo{pages}{67} (\bibinfo{year}{1994}).

\bibitem[{\citenamefont{Botet et~al.}(1982)\citenamefont{Botet, Jullien, and
  Pfeuty}}]{Botet1982}
\bibinfo{author}{\bibfnamefont{R.}~\bibnamefont{Botet}},
  \bibinfo{author}{\bibfnamefont{R.}~\bibnamefont{Jullien}}, \bibnamefont{and}
  \bibinfo{author}{\bibfnamefont{P.}~\bibnamefont{Pfeuty}},
  \bibinfo{journal}{Phys.~Rev.~Lett.} \textbf{\bibinfo{volume}{49}},
  \bibinfo{pages}{478} (\bibinfo{year}{1982}).

\bibitem[{\citenamefont{Cirac et~al.}(1998)\citenamefont{Cirac, Lewenstein,
  M{\o}lmer, and Zoller}}]{Cirac1998}
\bibinfo{author}{\bibfnamefont{J.~I.} \bibnamefont{Cirac}},
  \bibinfo{author}{\bibfnamefont{M.}~\bibnamefont{Lewenstein}},
  \bibinfo{author}{\bibfnamefont{K.}~\bibnamefont{M{\o}lmer}},
  \bibnamefont{and} \bibinfo{author}{\bibfnamefont{P.}~\bibnamefont{Zoller}},
  \bibinfo{journal}{Phys.~Rev.A} \textbf{\bibinfo{volume}{57}},
  \bibinfo{pages}{1208} (\bibinfo{year}{1998}).

\bibitem[{\citenamefont{You et~al.}(2006)\citenamefont{You, Wang, Tanamoto, and
  Nori}}]{You2006}
\bibinfo{author}{\bibfnamefont{J.~Q.} \bibnamefont{You}},
  \bibinfo{author}{\bibfnamefont{X.}~\bibnamefont{Wang}},
  \bibinfo{author}{\bibfnamefont{T.}~\bibnamefont{Tanamoto}}, \bibnamefont{and}
  \bibinfo{author}{\bibfnamefont{F.}~\bibnamefont{Nori}},
  \emph{\bibinfo{title}{Efficient one-step generation of large cluster states
  with solid-state circuits}} (\bibinfo{year}{2006}),
  \urlprefix\url{quant-ph/0609123}.

\bibitem[{\citenamefont{Roy and Braunstein}(2006)}]{Roy2006}
\bibinfo{author}{\bibfnamefont{S.~M.} \bibnamefont{Roy}} \bibnamefont{and}
  \bibinfo{author}{\bibfnamefont{S.~L.} \bibnamefont{Braunstein}},
  \emph{\bibinfo{title}{Exponentially enhanced quantum metrology}}
  (\bibinfo{year}{2006}), \urlprefix\url{quant-ph/0607152}.

\bibitem[{\citenamefont{Nielsen and Chuang}(2000)}]{Nielsen2000}
\bibinfo{author}{\bibfnamefont{M.~A.} \bibnamefont{Nielsen}} \bibnamefont{and}
  \bibinfo{author}{\bibfnamefont{I.~L.} \bibnamefont{Chuang}},
  \emph{\bibinfo{title}{Quantum Computation and Quantum Information}}
  (\bibinfo{publisher}{Cambridge University Press},
  \bibinfo{address}{Cambridge}, \bibinfo{year}{2000}).

\bibitem[{not({\natexlab{a}})}]{note1}
\bibinfo{note}{Letting $|M_L\rangle$ be the eigenvector of $L=H+K$ with maximum
  eigenvalue $M_L$, we have $M_L=\langle M_L|L|M_L\rangle=\langle
  M_L|H|M_L\rangle+\langle M_L|K|M_L\rangle\le M_H+M_K$, and similarly for
  $|n_L\rangle$. The triangle inequality follows.}

\bibitem[{not({\natexlab{b}})}]{note2}
\bibinfo{note}{Maximization of $\Delta^2 H$ can be carried out in two steps.
  First maximize $\Delta^2 H$ for fixed $\langle H\rangle=\mu M_H+(1-\mu)m_H$
  to give $\Delta^2 H=\mu(1-\mu)(M_H-m_H)^2$, corresponding to probabilities
  $p_M=\mu$ and $p_m=1-\mu$ for the maximum and minimum eigenvalues.
  Maximization over $\mu$ then gives the maximum variance $(M_H-m_H)^2/4$ at
  $\mu=1/2$.}

\bibitem[{\citenamefont{Geremia et~al.}(2003)\citenamefont{Geremia, Stockton,
  Doherty, and Mabuchi}}]{Geremia2003}
\bibinfo{author}{\bibfnamefont{J.}~\bibnamefont{Geremia}},
  \bibinfo{author}{\bibfnamefont{J.~K.} \bibnamefont{Stockton}},
  \bibinfo{author}{\bibfnamefont{A.~C.} \bibnamefont{Doherty}},
  \bibnamefont{and} \bibinfo{author}{\bibfnamefont{H.}~\bibnamefont{Mabuchi}},
  \bibinfo{journal}{Phys. Rev. Lett.} \textbf{\bibinfo{volume}{91}},
  \bibinfo{pages}{250801} (\bibinfo{year}{2003}).

\bibitem[{\citenamefont{Cornish et~al.}(2000)\citenamefont{Cornish, Claussen,
  Roberts, Cornell, and E.Wieman}}]{Cornish2000}
\bibinfo{author}{\bibfnamefont{S.~L.} \bibnamefont{Cornish}},
  \bibinfo{author}{\bibfnamefont{N.~R.} \bibnamefont{Claussen}},
  \bibinfo{author}{\bibfnamefont{J.~L.} \bibnamefont{Roberts}},
  \bibinfo{author}{\bibfnamefont{E.~A.} \bibnamefont{Cornell}},
  \bibnamefont{and} \bibinfo{author}{\bibfnamefont{C.}~\bibnamefont{E.Wieman}},
  \bibinfo{journal}{Phys.~Rev.~Lett.} \textbf{\bibinfo{volume}{85}},
  \bibinfo{pages}{1795} (\bibinfo{year}{2000}).

\end{thebibliography}
\end{document}